**All-optical Reconfiguration of Ultrafast Dichroism in Gold Metasurfaces**

*Andrea Schirato, Andrea Toma, Remo Proietti Zaccaria, Alessandro Alabastri, Giulio Cerullo, Giuseppe Della Valle\*, and Margherita Maiuri\**

A. Schirato, Prof. G. Cerullo, Prof. G. Della Valle, Prof. M. Maiuri
Dipartimento di Fisica, Politecnico di Milano, P.zza Leonardo da Vinci 32, 20133 Milan, Italy
E-mail: giuseppe.dellavalle@polimi.it, margherita.maiuri@polimi.it
A. Schirato, Prof. A. Toma, Prof. R. Proietti Zaccaria
Istituto Italiano di Tecnologia, via Morego 30, 16163 Genoa, Italy
Prof. Remo Proietti Zaccaria
Cixi Institute of Biomedical Engineering, Ningbo Institute of Materials Technology and Engineering, Chinese Academy of Sciences, Ningbo 315201, P.R. China
Prof. A. Alabastri
Department of Electrical and Computer Engineering, Rice University, 66100 Main Street, Houston, TX 77005, USA
Prof. G. Cerullo, Prof. G. Della Valle, Prof. M. Maiuri
Istituto di Fotonica e Nanotecnologie – Consiglio Nazionale delle Ricerche, P.zza Leonardo da Vinci 32, 20133 Milan, Italy

Keywords: reconfigurable metasurfaces, all-optical switching, ultrafast nanophotonics, pump-probe spectroscopy, photoinduced anisotropy

Optical metasurfaces have come into the spotlight as a promising platform for light manipulation at the nanoscale, including ultrafast all-optical control via excitation with femtosecond laser pulses. Recently, dichroic metasurfaces have been exploited to modulate the polarization state of light with unprecedented speed. Here, we theoretically predict and experimentally demonstrate by pump-probe spectroscopy the capability to reconfigure the ultrafast dichroic signal of a gold metasurface by simply acting on the polarization of the pump pulse, which is shown to reshape the spatio-temporal distribution of the optical perturbation. The photoinduced anisotropic response, driven by out-of-equilibrium carriers and extinguished in a sub-picosecond temporal window, is readily controlled in intensity by tuning the polarization direction of the excitation up to a full sign reversal. This work proves that nonlinear metasurfaces offer the flexibility to tailor their ultrafast optical response in a fully all-optically reconfigurable platform.

# 1. Introduction

With the aim of achieving control over the properties of light at the nanoscale, a great effort has been lately devoted to the design and engineering of optical nanostructures,[1-5] with a particular interest in ultrathin components, known as metasurfaces.[6-8] These are planar arrangements of sub-wavelength nanostructures (referred to as meta-atoms, being either dielectric or metal-based) which, thanks to their highly flexible design, have been shown to behave as exceptional tools for the manipulation of light, with no counterparts among naturally available materials.[9-13]

Interestingly, the potential of optical metasurfaces is not limited to the implementation of linear functionalities, such as wavefront shaping, beam steering, static polarization management (see e.g. ref.[14] and references therein). Indeed, such nanostructured materials are particularly desirable for nonlinear optical applications as well,[15-18] because of both the strong field enhancements they feature and the possibility they offer to modify in time their optical response via nonlinear effects. In particular, using intense femtosecond laser pulses,[19, 20] this modification takes place on ultrafast time scales, offering the unique opportunity to manipulate light by all-optical means with unprecedented speed and using ultracompact platforms.[21, 22]

In the plethora of configurations explored so far, plasmonic metasurfaces have taken on a notable significance in light of the giant nonlinearity dominating their optical response.[23-25] Indeed, upon photoexcitation with ultrashort pulses, a third-order delayed nonlinear process[26, 27] presides over the plasmon dephasing and subsequent generation of hot carriers.[28] In turn, the latter undergo internal relaxation processes[29] to equilibrate with the metal lattice and eventually with the environment by dissipating the excess energy deposited by the excitation pulse at ultrafast speed.

Among the diverse functionalities for which metasurfaces have been employed, all-optical control of the polarization state of light has been investigated with the goal to achieve sub-picosecond switching via modulation of the dichroism in the meta-atoms. In this framework, the schemes considered so far involved structures exhibiting either an intrinsic non-zero dichroic response,[30-34] or a transient photoinduced dichroism.[35, 36] However, the possibility to reconfigure this dichroism, dictated by the anisotropy (either geometrical or optical) of the structure, has not been explored so far, unless on extremely long time scales.[37, 38]

In this work, we theoretically predict and experimentally demonstrate all-optical reconfiguration of the ultrafast dichroism photogenerated in a nonlinear plasmonic metasurface consisting of symmetric gold cross-shaped meta-atoms.[35] We use polarization-controlled femtosecond pump-probe spectroscopy to measure the switch in sign of the photoinduced anisotropy (and subsequent dichroism), occurring on a sub-picosecond timescale and over a broad spectral region. This effect is governed by the spatio-temporal dynamics of the photogenerated hot carriers – a process almost overlooked until very recently[31, 35, 39-42] – which is intimately linked to the illumination conditions. In particular, we demonstrate the ability to reconfigure the sign of the ultrafast dichroism by rotating the polarization of the exciting pulse, a degree of freedom so far unexplored. Driven by technological interest towards multifunctional compact devices for ultrafast nanophotonics, our results provide evidence that nonlinear metasurfaces represent a simple yet robust platform for tailoring the properties of light, such as the polarization condition.

## 2. Results and Discussion

The rationale behind our study is sketched in **Figure 1**. We choose a nonlinear metasurface consisting of a periodic square lattice of gold cross-shaped nanoparticles (Figure 1a, refer to the Methods Section for details on the sample). Since the arms of the cross have equal length, the single meta-atom shows a four-fold symmetry (i.e., it is invariant under in-plane rotations by 90°), which is inherited in turn by the metamaterial. As a result, the latter exhibits an isotropic optical response when illuminated with linearly polarized radiation at normal incidence (Figure 1b). That is, the static transmission (*T*), absorption (*A*) and reflection (*R*) spectra, dominated by a plasmonic resonance peaked at ~800 nm (Figure 1b), are invariant under rotation of the polarization angle $\vartheta$, defined as the angle formed between the polarization direction of the incident beam and the vertical arm of the cross (as sketched in Figure 1c).

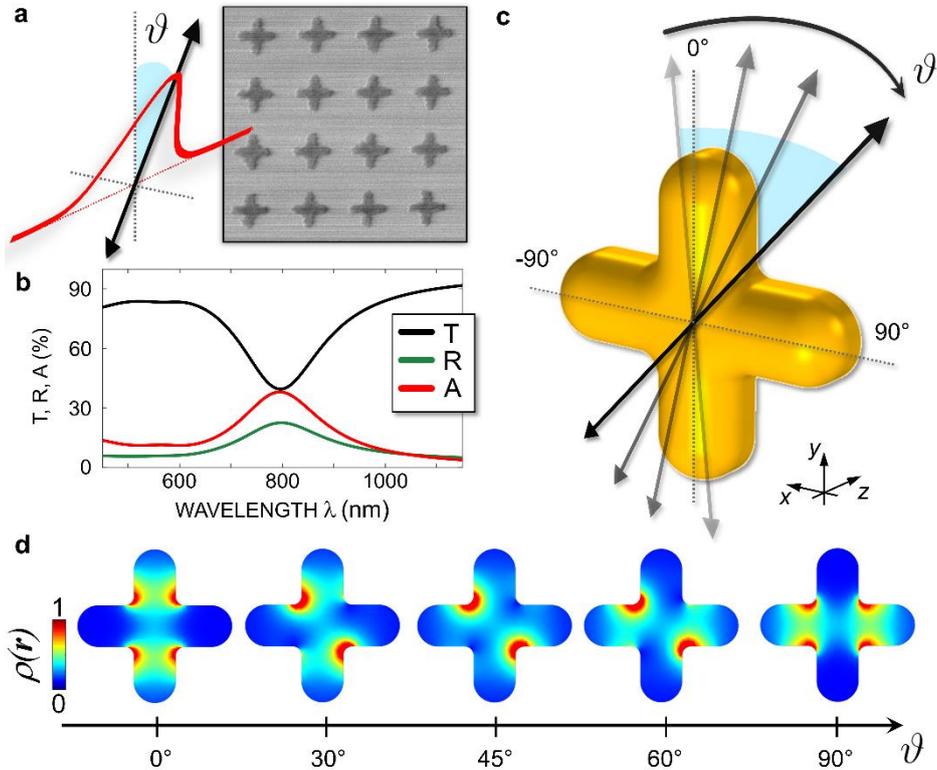

**Figure 1.** Control over optical symmetry via light polarization. a) Scanning Electron Microscope (SEM) image of the investigated plasmonic metasurface, consisting of a square lattice of Au symmetric nanocrosses. The system is photoexcited with light linearly polarized along a direction defining an angle $\vartheta$ with the (in-plane) vertical axis. b) Simulated static optical behaviour of the metasurface, detailed in terms of transmission (black), absorption (red), and reflection (green). In unperturbed conditions, the linear optical response of the structure is perfectly isotropic. c) Schematic of the light-matter interaction between the cross-shaped meta-atom and the linearly polarized light with varying polarization angles $\vartheta$. d) Normalized spatial pattern of the electromagnetic power density, $\rho(\mathbf{r})$, within the single metallic nanostructure evaluated at 860 nm for different polarization angles. Some exemplary conditions of photoexcitation are displayed across a horizontal (*xy*) cross-sectional plane at half-height of the meta-atom, corresponding to $\vartheta = 0°, 30°, 45°, 60°, 90°$.

Conversely, significant dependence on the polarization angle is observed for the near fields within the single meta-atoms. Locally, due to the relatively large dimensions of the nanocross (length, width and height respectively equal to 165 nm, 60 nm, 45 nm), the plasmonic modes excited by linearly polarized light are characterized by well-structured spatial patterns, changing substantially with $\vartheta$. As a result, the electromagnetic power density, which governs the absorption of light in the metal, inherits a polarization-dependent inhomogeneous distribution across the nanoparticle from the plasmon near fields. Figure 1d shows the (normalized) power dissipation density $\rho(\boldsymbol{r})$, with $\boldsymbol{r}$ the spatial coordinate in the meta-atom, at an exemplary wavelength of 860 nm for a set of polarization angles, ranging from 0° (polarization parallel to the cross vertical arm) to 90° (polarization orthogonal to the cross vertical arm). The crucial role of polarization direction manifests itself when the computed spatial patterns are compared. Indeed, the relative orientation of light polarization with respect to the nanocross axes determines a characteristic absorption pattern. The positions of the induced electromagnetic hot spots change with $\vartheta$, substantially following the direction of the oscillation gradient of the incoming electric field and identifying deeply sub-wavelength regions where the power is more concentrated.

The observed dependence of $\rho(\boldsymbol{r})$ on $\vartheta$ can be qualitatively interpreted by reducing the cross-shaped nanoparticle be to a superposition of two orthogonal and frequency degenerate plasmonic nanorods. At the leading order, when the polarization angle is either 0° or 90°, light mostly excites a plasmonic mode reminiscent of the longitudinal resonance of a single arm, either the vertical or the horizontal one. On the contrary, when light impinges with a polarization angle of 45°, the contributions of the two arms have equivalent weights and a critical, perfectly balanced configuration is achieved. Any angle between these conditions is expected to induce an imbalanced spatial dissipation pattern with intermediate symmetry.

In these terms, the angle of polarization $\vartheta$ becomes a crucial degree of freedom through which the symmetry of the dissipated power density can be effectively controlled. Importantly, the spatial distribution of $\rho(\mathbf{r})$ impacts the system optical response. Indeed, electromagnetic absorption governs changes of the electronic population in the metal, where plasmon-assisted photogeneration of hot carriers occurs. Electrons are promoted to out-of-equilibrium energetic states with a spatially non-uniform distribution determined by the local inhomogeneities of the absorption pattern. In this way, the hot carriers modify the metal permittivity non-uniformly in space and give rise to an ultrafast optical anisotropy at the nanoscale, which can be reconfigured by acting on the polarization angle $\vartheta$.

We use femtosecond pump-probe spectroscopy to characterize the all-optically reconfigurable dichroism. By tuning the direction $\vartheta$ of the pump pulse linear polarization, we control the symmetry of the spatial absorption pattern and the features of the transient nonlinear optical response. A delayed probe pulse, impinging at normal incidence with linear polarization at 45° to the nanocross arms, interrogates the optically excited metasurface. The measured signal $\Delta T/T$, as a function of the probe wavelength $\lambda$ and the pump-probe time delay $\tau$, is the difference between the probe spectra transmitted by the excited $T'(\lambda,\tau)$ and unexcited $T(\lambda)$ system, normalized to the unexcited probe transmittance $T(\lambda)$, $\Delta T/T = T'(\lambda,\tau)/T(\lambda) - 1$. To reveal the fingerprint of a $\vartheta$-controlled symmetry breaking, the $\Delta T/T$ is analyzed along the polarization directions corresponding to the vertical (0°) and horizontal (90°) arms of the nanocross (Figure 2a). This provides two distinct differential transmission signals, $\Delta T_{0°}(\lambda,\tau)/T(\lambda) = \Delta_0(\lambda,\tau)$ and $\Delta T_{90°}(\lambda,\tau)/T(\lambda) = \Delta_{90}(\lambda,\tau)$. The difference $\Delta_{90} - \Delta_0$ is thus a measure of the hot carrier-driven ultrafast anisotropy.[35] The dependence of the latter on the angular direction of the pump polarization should affect, instead, the intensity of $\Delta_{90} - \Delta_0$, over a broad range of wavelengths. Very simple arguments can be invoked to predict the trend of $\Delta_{90} - \Delta_0$ as function of the pump polarization angle $\vartheta$. Specifically, the variation of probe transmission for polarization direction parallel to the nanocross vertical arm, $\Delta T_{0°}(\lambda,\tau)$, is expected to be proportional to the projection of the pump along that direction, i.e. to $cos^2\vartheta$. On the other hand, when the transmitted probe intensity is detected along the direction of the nanocross horizontal arm, the resulting transient signal, $\Delta T_{90°}(\lambda,\tau)$, should be proportional to $sin^2\vartheta$. This suggests that the ensuing differential transient dichroism $\Delta_{90} - \Delta_0$ could be written as proportional to the difference $sin^2\vartheta - cos^2\vartheta \propto cos(2\vartheta)$, at any given pump-probe time delay and probe wavelength.

In fact, despite referring to a simplified model of the system, this straightforward reasoning is supported by a rigorous numerical model of the metasurface dynamical response. To this purpose, the Inhomogeneous Three-Temperature Model (I3TM),[35, 42] a freshly reported extension of the well-established 3TM,[43, 44] has been employed. Briefly, this is a semi-classical model describing the photogeneration and ultrafast relaxation of hot carriers in terms of three space-dependent energetic degrees of freedom within the plasmonic system. These are the excess energy stored in a non-thermalized portion of the out-of-equilibrium electrons, $N(r,t)$, the temperature of thermalized electrons, $\Theta_E(r,t)$, and the metal lattice temperature, $\Theta_L(r,t)$. The spatial dependences of those quantities are accounted for by including the spatial distribution in the drive term, dictated by the electromagnetic dissipation $\rho(r)$, and Fourier-like diffusion terms for the electron and phonon temperatures. The spatio-temporal dynamics of the three plasmonic degrees of freedom are then used to compute the corresponding changes of the metal permittivity, according to a widespread description of the thermo-modulational nonlinearities in Au.[45] Further details on the model are provided in the Methods Section.

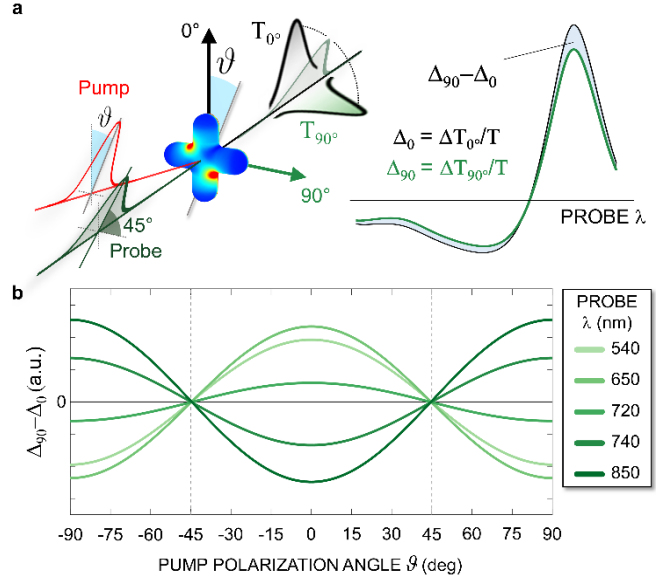

**Figure 2.** Polarization-selective pump-probe spectroscopy for angular-dependent transient dichroism. a) Sketch of the experimental set-up. For a fixed polarization angle $\vartheta$ of the pump pulse, meta-atoms experience an optical perturbation with a well-defined symmetry. Probe pulses interrogating the sample at a certain time delay from the pump with linear polarization at 45° to the nanocross arms are analyzed along directions at 0° (black arrow and pulse) and 90° (green arrow and pulse). Such decomposition provides two broadband differential transmission spectra, $\Delta_0$ (black solid) and $\Delta_{90}$ (green solid). The difference between these differential spectra (shaded in light blue) is the dichroic figure of merit. b) Predicted dichroic signal, $\Delta_{90} - \Delta_0$, for the metasurface of Au nanocrosses following photoexcitation with a pump pulse linearly polarized along $\vartheta$ direction. The broadband dichroism photoinduced by the pump is here displayed for a fixed pump-probe time delay $\tau = 100$ fs at a few selected probe wavelengths.

The results of the predicted dichroic quantity $\Delta_{90} - \Delta_0$ as a function of the pump polarization angle $\vartheta$ are reported in Figure 2b for a few exemplary probe wavelengths in the visible, at a fixed pump-probe delay $\tau = 100$ fs. The computed $\Delta T/T$ spectra were obtained for a pump at $\lambda_p = 860$ nm (at the red edge of the metasurface plasmonic peak, Figure 1b), by varying the pump polarization conditions.

As anticipated by our simple arguments on the transmitted signal angular projections, the $\Delta_{90} - \Delta_0$ dichroic signal follows a *cos(2$\vartheta$)*-like dependence precisely. Each wavelength features a different amplitude, since the light-induced dynamical modulation of Au permittivity is wavelength dependent, and so is the resulting differential transmittance. However, regardless of their absolute values, each curve in Figure 2b reaches its extrema in 0° and 90° and vanishes when $\vartheta = \pm 45°$. Such peculiar trend, related to the periodicity of the *cos(2ϑ)* function, highlights the many opportunities of controlling the ultrafast dichroism by photoexcitation. First, by changing $\vartheta$, following the curves of Figure 2b, the transient dichroic signal can be easily tuned: the highest value is achieved when the excited plasmonic mode is mostly concentrated in one of the nanocross arms (either 0° or 90°). Moreover, in the immediate proximity of these values, small angular variations only marginally affect the dichroism, providing the most robust excitation conditions (the signal derivative is the lowest). However, tuning $\vartheta$ does not only offer the possibility to change the magnitude of the effect. Even more interestingly, two pump pulses with polarization angles differing by 90° (or, equivalently, angles which are symmetric with respect to $\vartheta = \pm 45°$) allow for a full switch in the sign of the induced dichroism. Therefore, in the region around ±45°, the structure is particularly sensitive to variations of the polarization orientation, and $\Delta_{90} - \Delta_0$ undergoes a sign reversal at any wavelength when pairs of angles differing by 90° are compared.

In order to prove the actual possibility to achieve the predicted control over the sign of the transient dichroism via fine-tuning of the polarization angle of an ultrashort pump pulse, polarization-resolved pump-probe measurements have been performed over a broad range of visible wavelengths at two distinct excitation conditions (refer to the Methods Section for details on the experimental set-up). The ultrafast experiments have then been combined with the numerical simulations introduced before.

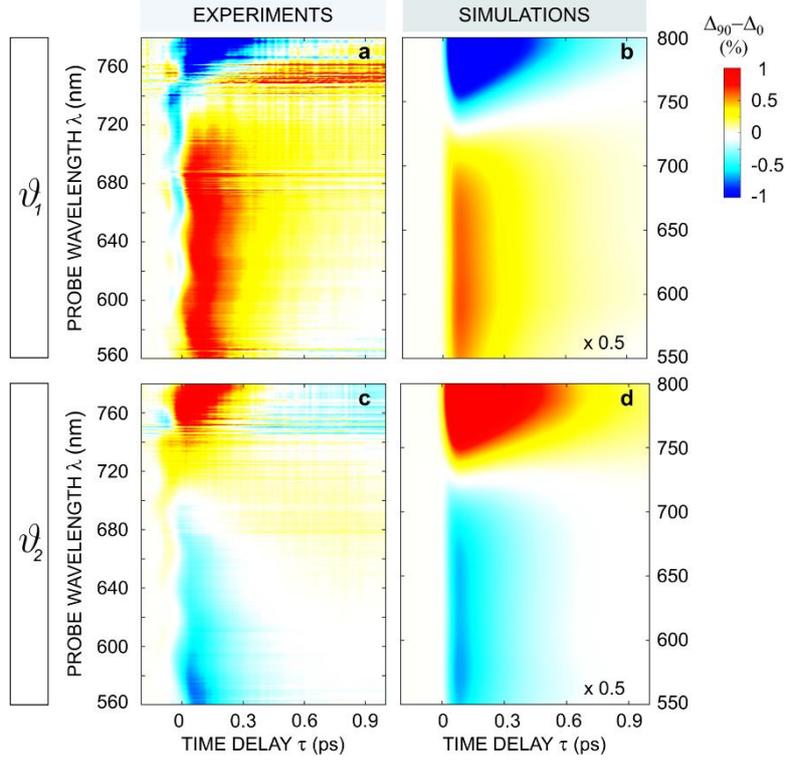

**Figure 3.** Sign reversal of photoinduced ultrafast dichroism. a - d) Experimental (a, c) and simulated (b, d) pump-probe maps of the broadband transient dichroism, $\Delta_{90} - \Delta_0$, photoinduced in the plasmonic metasurface by pump pulse absorption (pump wavelength $\lambda_p$ = 860 nm, fluence Fp ~ 400 µJ cm$^{-2}$) for two polarization angles $\vartheta_1$ and $\vartheta_2 = \vartheta_1 + 90°$. Such choice allows for reversing the sign of the dichroic figure over the entire broad range of wavelengths analyzed.

A first set of measurements has been performed with a pump polarization angle $\vartheta_1 = 40° \pm 2°$ (Figure 3a). In modelling the experimental results, an angle $\vartheta_1 = 38°$ has been considered to match with the measurements (Figure 3b), which is within the experimental accuracy of determination of $\vartheta$. The obtained dichroism (Figure 3 top panels) exhibits temporal and spectral features in substantial agreement with previous results.[35] $\Delta_{90} - \Delta_0$ remains positive and spectrally almost flat over a broad range of wavelengths, from 560 nm to 720 nm, before changing sign at ~730 nm, where a well-defined negative band is observed at longer wavelengths. The photoinduced dichroism is maximum at ~100 fs, when it reaches ~1%. Note that this is the same order of magnitude as the differential transmittance recorded for the two polarizations, although lower than the dichroism reported in ref.[35], obtained in the optimal (and the most robust) conditions for symmetry-breaking, i.e., pump pulse polarization aligned to one of the cross arms. Then, at longer time delays, the signal of $\Delta_{90} - \Delta_0$ decays much faster than the differential transmission signal, extinguishing in less than 1 ps, as it is governed by the homogenization processes of the hot carrier population within the nanocross, which recovers an excited yet isotropic state within a few hundreds of femtoseconds. Simulations well reproduce the experimental measurements, apart from a slight spectral shift and an intensity scaling factor. The latter is expected to be due to the limits of the semiclassical calculation of non-thermal carrier photogeneration.[35,46,47] Likewise, experiments and simulations are repeated by rotating the pump polarization direction by 90°, providing the dichroism maps of Figure 3c, 3d (bottom panels) at $\vartheta_2 = \vartheta_1 + 90°$. Results manifestly demonstrate the predicted sign reversal with the polarization angle tuning. In the two excitation conditions, the ultrafast dichroism shows comparable features, but it experiences a clear switch in sign over the entire spectrum (compare positive and negative bands in Figure 3a and 3c for experiments, 3b and 3d for simulations). In other words, the ultrafast broadband response is utterly switched from positive to negative and vice-versa within the same structure by simple rotation of pump light polarization.

To better visualize the broadband sign switch, Figure 4 shows selected spectral cuts of the ΔT/T maps. In particular, spectra of the measured dichroism for $\vartheta_1$ and $\vartheta_2$ (refer to Methods Section for details) at a pump-probe delay of 100 fs are reported in Figure 4a and 4b respectively. The same comparison is shown for simulated signals (Figure 4c and 4d).

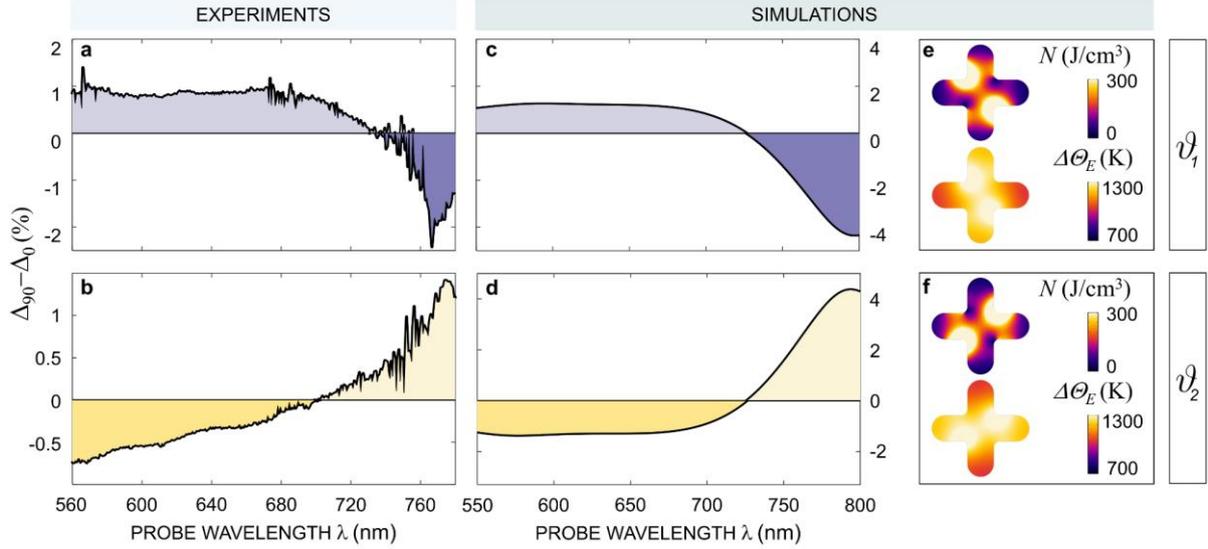

**Figure 4.** Spectra of the transient dichroism. a - d) Experimental (a, b) and simulated (c, d) spectra of the dichroism, $\Delta_{90} - \Delta_0$, evaluated at a pump-probe time delay $\tau = 100$ fs. Top panels (a, c) refer to the excitation condition with pump polarization angle $\vartheta_1$, bottom panels (b, d) are obtained when the pump pulse is polarized along the $\vartheta_2$ direction. e - f) Spatial distribution of the energy density, $N$, stored in the non-thermalized portion of hot carriers (top) together with the electronic temperature increase $\Delta\Theta_E$ (bottom) across the central cross-section (in the $x$ direction) of the single meta-atom, evaluated at $\tau = 100$ fs, when the pump pulse polarization angle is either $\vartheta_1$ (e) or $\vartheta_2$ (f).

The spectra of the photoinduced dichroism clearly display the reversal of signals: both bands switch in sign when moving from $\vartheta_1$ (top panels) to $\vartheta_2$ (bottom panels), with a rather good agreement between experiments and simulations. Note that the zero-crossing points of spectra measured for the two polarization angles (Figure 4a, 4b) do not fall precisely at the same wavelengths (that is, spectra are not simply flipped), and isosbestic points of the two spectra follow slightly different temporal evolution. A similar trend, albeit to a lesser degree, is also observed in simulations (compare Figure 3b, 3d, and Figure 4c, 4d) and can be explained by the small (~5°) angle formed by the wave-vector of the pump pulse with the normal to the sample plane. This angle, slightly tilting the symmetry of the optical axes photoinduced in the nanocross, has been proved to lightly affect the symmetry-breaking mechanism.[35]

Moreover, simulations also disclose the fundamental connection between the measured signal and hot carriers at the nanoscale (Figure 4e and 4f). Inspecting $N$ and $\Delta\Theta_E$ across the single meta-atoms (Figure 4e for $\vartheta_1$, 4f for $\vartheta_2$, both referring to a pump-probe delay $\tau = 100$ fs) suggests an explanation for the sign reversal in terms of the symmetry of their spatial distributions. The non-uniform patterns of the high energy electrons change according to the excitation condition and, when compared, are substantially symmetric with respect to a rotation of $\vartheta_2 - \vartheta_1 = 90°$. Mirrored inhomogeneities produce opposite changes in the optical properties along the co-orthogonal directions probed (0° and 90°), thus promoting the sign switch. In these terms, the photoinduced dichroism and its full reconfiguration are demonstrated to be substantially governed by the ultrafast spatial dynamics of hot electrons at the nanoscale

**3. Conclusion**

In this study, we predicted and simulated a sign reversal of the ultrafast optical response of a plasmonic metasurface, exhibiting transient dichroism modulated by varying the excitation conditions. The observed switch in sign is shown to intrinsically rely on fundamental properties of light absorption in nanostructures and the spatio-temporal evolution of photogenerated hot carriers.

Our investigation focussed on a quasi-2D metamaterial consisting of (four-fold) symmetric cross-shaped meta-atoms, characterized by an isotropic optical behaviour in unperturbed conditions. Ultrafast pump-probe spectroscopy was used to reveal a transient anisotropy in the structure, previously reported to be governed by the local spatial inhomogeneities of photogenerated hot carriers across the single meta-atoms. Indeed, pump absorption promotes an out-of-equilibrium and spatially inhomogeneous electron excitation, which in turn drives a spatially non-uniform permittivity modulation in the metal and results in a transient dichroic response. Importantly, the spatial distribution of the hot-carrier population is ultimately dictated by the absorption pattern of the exciting pulse, which establishes the symmetry of the perturbation. We can therefore achieve full reconfiguration of the metasurface response by simply tuning the polarization direction of the pump pulse. In particular, a sign reversal of the dichroism was experimentally demonstrated for a pair of pump pulses with polarizations differing by 90°. Indeed, the relative angle condition induces opposite symmetries for the transient anisotropy, thus offering the flexibility to tailor the sign of the optical response. We also proposed simple arguments to explain the signal modulation in terms of projections of the optical perturbation experienced by probe pulses interrogating the metasurface, and we reproduced the experiments via semiclassical model of the Au nonlinearities and the nanocross electromagnetic behaviour.

Our experimental and numerical results show how polarization-selective light absorption in plasmonic metasurfaces and the resulting hot carrier deep sub-wavelength inhomogeneities allow for active shaping of the ultrafast response, with no need to change the geometrical design. Furthermore, the simplicity of our approach in attaining ultrafast reconfiguration of dichroism suggests that a similar rationale may be applied to envisage wide-ranging flexible systems interleaving various functionalities on ultrafast time scales within a single all-optical device.

## 4. Methods

*Sample Fabrication*: The sample consists of a metasurface made of closely packed cross-shaped nanoparticles fabricated by electron beam lithography. Meta-atoms are arranged in a square lattice with centre-to-centre nominal distance of 270 nm. The metasurface is supported on a substrate made of $CaF_2$. Further details on the fabrication process of the sample can be found in ref.[35]

*Ultrafast Pump-Probe Spectroscopy*: A detailed description of the apparatus used for the ultrafast pump-probe spectroscopy is provided elsewhere.[48] Briefly, an amplified Ti:sapphire laser, operating at 2-kHz repetition rate, is used to pump two non-collinear optical parametric amplifiers, each of which generates the pump and probe pulses respectively. Pump pulses span over the 850–950-nm spectral region and were compressed by a fused-silica prism pair to sub-30-fs duration, while probe pulses cover the 560–780-nm bandwidth and were compressed by chirped mirrors to sub-10-fs duration. The pump and probe were focused on the sample by a spherical mirror to a diameter of 50 μm. The transmitted probe was dispersed in a spectrometer and detected with a custom-made charge-coupled device operating at the full laser repetition rate. The metasurface was aligned perpendicular to the probe beam.

To measure the sign reversal of the transient dichroism, we set the pump polarization at two fixed angles $\vartheta_1$ and $\vartheta_2$ (with $\vartheta$ defined as in Figure 1 by the polarization direction with respect to the vertical arms of the crosses), given by $\vartheta_1 = 40° \pm 2°$ and $\vartheta_2 = \vartheta_1 + 90°$, while the probe polarization was fixed at an angle of 45° with respect to the nanocross arm. The measured differential transmittance, for each pump polarization angle $\vartheta$, was analyzed along two co-orthogonal directions corresponding to the vertical (0°) and horizontal (90°) cross arms, thus providing two distinct pump-probe signals, $\Delta T_{90°,0°}(\lambda,\tau)/T(\lambda) = \Delta_{90,0}(\lambda,\tau)$ (see Figure 2a). Their difference $\Delta_{90} - \Delta_0$ represents the measured figure of merit related to the photogenerated ultrafast anisotropy, as a function of probe wavelength and pump-probe delay.

*Numerical Modelling*: To describe the ultrafast optical response of the plasmonic metasurface investigated, we employed a multi-step numerical simulation and pursued a semi-classical modelling approach evaluating the transient transmittance modulation of the structure. The starting point of our model is the description of photogeneration and relaxation of hot carriers. This was done by means of the I3TM[35, 42], which accounts for the evolution both in time and in space of hot electrons in gold meta-atoms upon illumination with an ultrashort laser pulse. Specifically, a segregated algorithm was implemented in the Finite Element Method (FEM)-based commercial software COMSOL Multiphysics, integrating the rate equations written for the plasmonic system energetic degrees of freedom[37,38]:

$$\frac{\partial N}{\partial t} = -aN - bN + P_{abs}(\boldsymbol{r},t) \qquad (1)$$

$$C_E \frac{\partial \Theta_E}{\partial t} = -\nabla(-\kappa_E \nabla \Theta_E) - G(\Theta_E - \Theta_L) + aN \quad (2)$$

$$C_L \frac{\partial \Theta_L}{\partial t} = \kappa_L \nabla^2 \Theta_L + G(\Theta_E - \Theta_L) + bN \quad (3)$$

where the equation coefficients govern the energy relaxation processes undergone by $N(r,t)$, $\Theta_E$ $(r,t)$ and $\Theta_L$ $(r,t)$ (details on their definitions and values can be found in ref.[35] and references therein), while the driving term of the excitation, $P_{abs}(r,t)$, represents the instantaneous density of electromagnetic power absorbed by the structure. Its expression, evolving in time with the Gaussian intensity profile of the pump pulse (modelled with a full-width at half-maximum time duration of 60 fs), is proportional to the absorption pattern introduced in the main text, $\rho(r)$. To evaluate the spatial distribution of the latter across the meta-atom, we employed 3D full-wave electromagnetic simulations in the frequency domain. FEM port formalism (employed to mimic the response of nanocrosses in periodic array configuration) allowed us to account for the dependence of $\rho(r)$ on the polarization condition of the pump, by adjusting the orientation of the incoming electric field.

Then, based on the evolution in time and space of the nanostructure internal energy variables, we determined the corresponding permittivity modulation, locally evolving in each metallic nanostructure. Spectrally dispersed transient changes of both the inter- and intraband terms of Au permittivity were modelled, following a semi-classical description of the third-order delayed optical nonlinearities in noble metals.[37,39] Indeed, plasmon-assisted photoexcitation of out-of-equilibrium carriers affects the occupation probability of the energy levels in the conduction band. This entails an intrinsically nonlinear variation of the optical absorption coefficient or, equivalently, the imaginary part of permittivity (the real part being readily computed by Kramers-Kronig analysis). In this framework, the additional dependence on space introduced by the I3TM for the energetic degrees of freedom ($N$, $\Theta_E$ and $\Theta_L$) is naturally included in the modelling approach[35] as the key ingredient to account for the ultrafast optical anisotropy following pump absorption. To reveal the obtained spatially inhomogeneous distribution of permittivity, the perturbed structure is interrogated over a broad spectral range by scanning the time delay between pump and probe. Full-wave frequency-domain simulations are employed, by suitably defining the numerical ports so to analyze the dynamical transmittance over the directions of interest, similarly to experimental conditions.

**Acknowledgements**
G.D.V. and M.M. contributed equally to this work.
This publication is part of the METAFAST project that received funding from the European Union Horizon 2020 Research and Innovation programme under Grant Agreement No. 899673. This work reflects only author view and the European Commission is not responsible for any use that may be made of the information it contains.

Received: ((will be filled in by the editorial staff))
Revised: ((will be filled in by the editorial staff))
Published online: ((will be filled in by the editorial staff))

References

[1] S. A. Maier, M. L. Brongersma, P. G. Kik, S. Meltzer, A. A. G. Requicha, H. A. Atwater, Advanced Materials 2001, 13, 1501.

[2] J. A. Schuller, E. S. Barnard, W. Cai, Y. C. Jun, J. S. White, M. L. Brongersma, Nature Materials 2010, 9, 193.

[3] M. L. Brongersma, Nature Photonics 2008, 2, 270.

[4] H. Fischer, O. J. F. Martin, Opt. Express 2008, 16, 9144.

[5] A. I. Kuznetsov, A. E. Miroshnichenko, M. L. Brongersma, Y. S. Kivshar, B. Luk'yanchuk, Science 2016, 354, aag2472.

[6] J. B. Pendry, D. Schurig, D. R. Smith, Science 2006, 312, 1780.

[7] C. M. Soukoulis, M. Wegener, Science 2010, 330, 1633.

[8] D. Neshev, I. Aharonovich, Light: Science & Applications 2018, 7, 58.

[9] A. V. Kildishev, A. Boltasseva, V. M. Shalaev, Science 2013, 339, 1232009.

[10] N. Yu, F. Capasso, Nature Materials 2014, 13, 139.

[11]     T. Phan, D. Sell, E. W. Wang, S. Doshay, K. Edee, J. Yang, J. A. Fan, Light: Science & Applications 2019, 8, 48.

[12]     N. I. Zheludev, Y. S. Kivshar, Nature Materials 2012, 11, 917.

[13]     S. B. Glybovski, S. A. Tretyakov, P. A. Belov, Y. S. Kivshar, C. R. Simovski, Physics Reports 2016, 634, 1.

[14]     H.-T. Chen, A. J. Taylor, N. Yu, Reports on Progress in Physics 2016, 79, 076401.

[15]     M. Ren, W. Cai, J. Xu, Advanced Materials 2020, 32, 1806317.

[16]     G. Li, S. Zhang, T. Zentgraf, Nature Reviews Materials 2017, 2, 17010.

[17]     J. Lee, M. Tymchenko, C. Argyropoulos, P.-Y. Chen, F. Lu, F. Demmerle, G. Boehm, M.-C. Amann, A. Alù, M. A. Belkin, Nature 2014, 511, 65.

[18]     A. Krasnok, M. Tymchenko, A. Alù, Materials Today 2018, 21, 8.

[19]     P. Vasa, C. Ropers, R. Pomraenke, C. Lienau, Laser & Photonics Reviews 2009, 3, 483.

[20]     S. V. Makarov, A. S. Zalogina, M. Tajik, D. A. Zuev, M. V. Rybin, A. A. Kuchmizhak, S. Juodkazis, Y. Kivshar, Laser & Photonics Reviews 2017, 11, 1700108.

[21]     G. A. Wurtz, R. Pollard, W. Hendren, G. P. Wiederrecht, D. J. Gosztola, V. A. Podolskiy, A. V. Zayats, Nature Nanotechnology 2011, 6, 107.

[22]     G. Della Valle, B. Hopkins, L. Ganzer, T. Stoll, M. Rahmani, S. Longhi, Y. S. Kivshar, C. De Angelis, D. N. Neshev, G. Cerullo, ACS Photonics 2017, 4, 2129.

[23]     M. Kauranen, A. V. Zayats, Nature Photonics 2012, 6, 737.

[24]     H. Baida, D. Mongin, D. Christofilos, G. Bachelier, A. Crut, P. Maioli, N. Del Fatti, F. Vallée, Physical Review Letters 2011, 107, 057402.

[25]     X. Wang, Y. Guillet, P. R. Selvakannan, H. Remita, B. Palpant, The Journal of Physical Chemistry C 2015, 119, 7416.

[26]     M. Conforti, G. Della Valle, Physical Review B 2012, 85, 245423.

[27] Y. He, W. Yang, T.-M. Shih, J. Wang, D. Zhang, M. Gao, F. Jiao, Y. Zeng, J.-L. Yang, J. Pang, R. Gao, G. Sun, M.-D. Li, J.-F. Li, Z. Yang, Advanced Optical Materials 2021, n/a, 2100847.

[28] A. Manjavacas, J. G. Liu, V. Kulkarni, P. Nordlander, ACS Nano 2014, 8, 7630.

[29] J. G. Liu, H. Zhang, S. Link, P. Nordlander, ACS Photonics 2018, 5, 2584.

[30] L. H. Nicholls, F. J. Rodríguez-Fortuño, M. E. Nasir, R. M. Córdova-Castro, N. Olivier, G. A. Wurtz, A. V. Zayats, Nature Photonics 2017, 11, 628.

[31] L. H. Nicholls, T. Stefaniuk, M. E. Nasir, F. J. Rodríguez-Fortuño, G. A. Wurtz, A. V. Zayats, Nature Communications 2019, 10, 2967.

[32] M. I. Shalaev, J. Sun, A. Tsukernik, A. Pandey, K. Nikolskiy, N. M. Litchinitser, Nano Letters 2015, 15, 6261.

[33] K. Wang, M. Li, H.-H. Hsiao, F. Zhang, M. Seidel, A.-Y. Liu, J. Chen, E. Devaux, C. Genet, T. Ebbesen, ACS Photonics 2021, 8, 2791.

[34] M. Taghinejad, H. Taghinejad, Z. Xu, K.-T. Lee, S. P. Rodrigues, J. Yan, A. Adibi, T. Lian, W. Cai, Nano Letters 2018, 18, 5544.

[35] A. Schirato, M. Maiuri, A. Toma, S. Fugattini, R. Proietti Zaccaria, P. Laporta, P. Nordlander, G. Cerullo, A. Alabastri, G. Della Valle, Nature Photonics 2020, 14, 723.

[36] A. Sugita, H. Yogo, K. Mochizuki, S. Hamada, H. Matsui, A. Ono, W. Inami, Y. Kawata, M. Yoshizawa, OSA Continuum 2020, 3, 2943.

[37] C. Ding, G. Rui, B. Gu, Q. Zhan, Y. Cui, Opt. Lett. 2021, 46, 2525.

[38] X.-T. Kong, L. Khosravi Khorashad, Z. Wang, A. O. Govorov, Nano Letters 2018, 18, 2001.

[39] A. Block, M. Liebel, R. Yu, M. Spector, Y. Sivan, F. J. García de Abajo, N. F. van Hulst, Science Advances 2019, 5, eaav8965.

[40] A. Rudenko, K. Ladutenko, S. Makarov, T. E. Itina, Advanced Optical Materials 2018, 6, 1701153.

[41]    Y. Sivan, M. Spector, ACS Photonics 2020, 7, 1271.

[42]    A. Schirato, A. Mazzanti, R. Proietti Zaccaria, P. Nordlander, A. Alabastri, G. Della Valle, Nano Letters 2021, 21, 1345.

[43]    M. Zavelani-Rossi, D. Polli, S. Kochtcheev, A.-L. Baudrion, J. Béal, V. Kumar, E. Molotokaite, M. Marangoni, S. Longhi, G. Cerullo, P.-M. Adam, G. Della Valle, ACS Photonics 2015, 2, 521.

[44]    C. K. Sun, F. Vallée, L. H. Acioli, E. P. Ippen, J. G. Fujimoto, Physical Review B 1994, 50, 15337.

[45]    R. Rosei, F. Antonangeli, U. M. Grassano, Surface Science 1973, 37, 689.

[46]    L. V. Besteiro, X.-T. Kong, Z. Wang, G. Hartland, A. O. Govorov, ACS Photonics 2017, 4, 2759.

[47]    L. V. Besteiro, P. Yu, Z. Wang, A. W. Holleitner, G. V. Hartland, G. P. Wiederrecht, A. O. Govorov, Nano Today 2019, 27, 120.

[48]    D. Polli, L. Lüer, G. Cerullo, Review of Scientific Instruments 2007, 78, 103108.

A combination of semiclassical modelling and pump-probe spectroscopy is exploited to demonstrate the all-optical reconfiguration of the ultrafast nonlinear response in a plasmonic metasurface made of gold nanocrosses. We show that the pump polarization tuning can reshape the spatio-temporal distribution of the photoinduced hot carriers to modulate the sub-picosecond anisotropy up to a full sign reversal.

Andrea Schirato, Andrea Toma, Remo Proietti Zaccaria, Alessandro Alabastri, Giulio Cerullo, Giuseppe Della Valle*, Margherita Maiuri*

**All-optical Reconfiguration of Ultrafast Dichroism in Gold Metasurfaces**

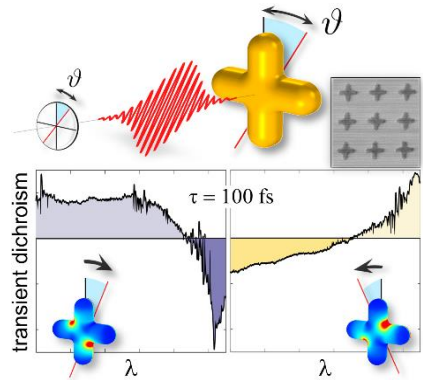